# Speeding up Simplification of Polygonal Curves using Nested Approximations

Pierre-François Marteau, Gildas Ménier

*VALORIA, Université de Bretagne Sud, BP573, 56017 Vannes France*
*Phone. : (33) 2 01 72 99,  Fax : (33) 2 01 72 79*
{pierre-francois.marteau, gildas.menier}@univ-ubs.fr, http://www-valoria.univ-ubs.fr/

**Abstract**: We develop a multiresolution approach to the problem of polygonal curve approximation. We show theoretically and experimentally that, if the simplification algorithm *A* used between any two successive levels of resolution satisfies some conditions, the multiresolution algorithm *MR* will have a complexity lower than the complexity of *A*. In particular, we show that if *A* has a $O(N^2/K)$ complexity (the complexity of a reduced search dynamic solution approach), where *N* and *K* are respectively the initial and the final number of segments, the complexity of *MR* is in $O(N)$. We experimentally compare the outcomes of *MR* with those of the optimal "full search" dynamic programming solution and of classical merge and split approaches. The experimental evaluations confirm the theoretical derivations and show that the proposed approach evaluated on *2D* coastal maps either shows a lower complexity or provides polygonal approximations closer to the initial curves.

## 1. Introduction

Approximation of multi dimensional discrete curves has been widely studied essentially to speed up data processing required by resource demanding applications such as Computer Vision, Computer Graphics, Geographic Information Systems and Digital Cartography, Data Compression or Time Series Data Mining. For polygonal approximation of discrete curves, the problem can be informally stated as follows: given a digitized curve *X* of $N \geq 2$ ordered samples, find *K* dominant samples among them that define a sequence of connected segments which most closely approximates the original curve. This problem is known as the *min-ε problem*. Numerous algorithms have been proposed for more than thirty years to solve efficiently this optimization problem. Most of them belong either to graph-theoretic approaches (Imai and Iri, 1986, 1988; Melkman and O'Rourke, 1988; Chan and Chin, 1996; Zhu, and Seneviratne, 1997; Chen and Daescu, 1998; Katsaggelos et al., 1998), dynamic programming (Perez and Vidal, 1994; Salotti, 2001; Horng, 2002, Kolesnikov et al., 2003, 2004) or to heuristic approaches (Douglas & Peucker, 1973; Pratt & Fink, 2002; Debled-Rennesson and Reveillès, 2003; Charbonnier al., 2004). This paper focuses on polygonal approximation of multidimensional curves using multiresolution. Our concern is the design of an algorithm that, starting from the finest resolution level, finds iteratively *min-$\varepsilon$* polygonal approximations for coarser resolution levels, exploiting at the any level the approximating segments obtained at the previous (finer) resolution level. In theory, such multiresolution approximation can exploit any polygonal simplification methods between two successive levels of resolution. Nevertheless, the effectiveness and efficiency of such approach is not guaranteed. The aim of this paper is to study the conditions required for efficiency and effectiveness of the multiresolution approach.

The first part of the paper addresses a complexity analysis that sketches conditions for the efficiency of a "top-down" multiresolution approach. Some argumentation is also provided to outline the conditions for effectiveness. A proposal for a simple multiresolution algorithm that fulfils the previous conditions is then proposed.

The second part of the paper addresses the evaluation of the proposed algorithm through experimentations that compare the efficiency and effectiveness of the multiresolution approach with heuristic and dynamic programming solutions.

## 2. A Multi Resolution approach to Polygonal Curve Approximation (*MR algorithm*)

Basically, the idea behind the multiresolution approach to polygonal curve simplification is to successively re-approximate previous approximations obtained by using some given simplification algorithm, this process being initiated from an original discrete time series. Following (Kolesnikov & al. 2004), we take a sequence of polygonal curves {$X_0$, $X_1$, $X_2$,..., $X_r$} as a multiresolution (multiscale) approximation of a *N*-vertex input curve *X*, if the set of curves {$X_i$} satisfies the following conditions:

i) A polygonal curve $X_i$ is an approximation of the curve *X* for a given number of segments $K_i$ (*min-ε problem*) or error tolerance $\varepsilon_i$ (*min-# problem*), where *i* is a resolution level (i=0,1,2,…, *r*).

ii) The set of vertices of curve $X_i$ for resolution level *i* is a subset of vertices of curve $X_{i-1}$ for the previous (higher) resolution level (*i-1*). The lowest resolution level *r* is represented by the coarsest approximation of *X*. The highest resolution level *i=0* is represented by the most detailed approximation (namely the original curve $X_0=X$) with the largest number of segments $K_0 =N$. ($N=K_0 > K_1 > K_2 >…> K_r$) or smallest error tolerance $\varepsilon_0=0$ for some distance measure (e.g. $L_2$) ($\varepsilon_0<\varepsilon_1<\varepsilon_2<…<\varepsilon_r$).

Thus, an approximation curve $X_i$ is either obtained by inserting new points into the approximation curve $X_{i+1}$ (bottom-up approach), or, conversely, $X_{i+1}$ is obtained by deleting points from the approximation curve $X_i$ (top-down approach). These two variants have led to the development of two very popular heuristic approaches: respectively *SPLIT* and *MERGE* methods. In the *SPLIT* approach, an iterative mechanism splits the input curve into smaller and smaller segments until the maximum deviation is smaller than a given error tolerance ε (*min-# problem*), or the number of linear segments equals to the given $K_i$ (*min-ε problem*) for the current resolution level *i*. A famous *SPLIT* method is the Douglas-Peucker algorithm (Douglas and Peucker, 1973); this algorithm is known to have a $O(K \cdot N)$ complexity; it has been used for multiresolution approximation in (Le Buhan Jordan & al., 1998, Buttenfield, 2002) and in (Kolesnikov & al. 2004) that developed the optimal split algorithm (*OSA*).

In the *MERGE* approach (Pikaz and Dinstein, 1995, Visvalingam and Whyatt 1993), the polygonal approximation is performed by using a cost function that determines sequential elimination of the vertices with the smallest cost value, while the two adjacent segments of the eliminated vertex are merged into one segment. The approximation curve $X_i$ is obtained by discarding vertices from the curve $X_{i-1}$ until the desired number of vertices $K_i$ (*min-ε problem*) is reached. This *Merge* approach is known to have a $O(N \cdot log(N))$ complexity.

## 2.1. Complexity of multiresolution *approach*

We address in this section the efficiency issue of the multiresolution approach comparatively to the complexity of an algorithm that provides the solution in a single step. Let $T$ be the complexity of an algorithm $A$ that simplifies a polygonal curve $X$ having $N$ segments into a polygonal curve having $K$ segments ($K<N$). We suppose that $T$ is a polynomial function of $N$ and $K$, namely $T = O(N^p K^q)$ where $p$ and $q$ are integers. For all $N$, $0<K<N$ and $\rho$ in $]0;1[$ there exists a natural number $r$ such that $N\rho^{r+1} < K \leq N\rho^r$. Given $N$, $K<N$ and $\rho$, we construct the multiresolution $\{X_0, X_1, X_2,..., X_r\}$ of curve $X$ such that $K_j = \rho^j N$ is the number of segments that compose $X_j$, the $j^{th}$ nested polygonal approximation of $X$.

In this context, using algorithm $A$ during the first step, we obtain from an original curve ($X$) of size $N$ a polygonal curve approximation ($X_1$) having $K_1 = \rho N$ segments with complexity: $T_1 = O(N^p K_1^q) = O(N^p (\rho\rho N^q) = O(\rho^q N^{p+q})$

Considering as a second step the simplification of the $X_1$ polygonal curve seen as a polygonal curve having $N_1=K_1$ samples, still using algorithm $A$ , we get a polygonal curve approximation $X_2$ having $K_2 = \rho^2 N$ segments from the $X_1$ curve with time complexity: $T_2 = O((K_1)^p (K_2)^q) = O((\rho(({}^p(\rho^2 N)^q) = O(\rho^{p+2q} N^{p+q})$

Iterating the process we get successively:

$$T_3 = O((K_2)^p (K_3)^q) = O((\rho^2 N)^p (\rho^3 N)^q) = O(\rho^{2p+3q} N^{p+q})$$

By induction, it is easy to show that for all $j$ in $\{1,..,r\}$ we have :

$$T_j = O((K_{j-1})^p (K_j)^q) = O((\rho^{j-1} N)^p (\rho^j N)^q) = O(\rho^{(j-1)p+jq} N^{p+q})$$

The final iteration - required to ensure that the last approximation has exactly $K$ segments – is performed using algorithm $\mathcal{A}$ with the following complexity:

$$T_{r+1} = O(K_r^p K^q) = O((\rho^r N)^p (K)^q)$$
as $\rho^{r+1} N \leq K < \rho^r N$, $\quad T_{r+1} \leq O((\rho^r N)^p (\rho^r N)^q) = O(\rho^{r(p+q)} N^{p+q})$ if $q > 0$,
$$T_{r+1} \leq O((\rho^r N)^p (\rho^{r+1} N)^q) = O(\rho^q N^{p+q} \rho^{r(p+q)}) \text{ otherwise.}$$

Finally, from the original time series of size $N$, we get after $r+1$ iterations of the previous process a polygonal approximation having $K$ segments with time complexity:

$$T_{MR} = T_{r+1} + \sum_{j=1}^{r} O(\rho^{(j-1)p+jq} N^{p+q}) = T_{r+1} + O(\rho^q N^{p+q} \sum_{j=1}^{r} \rho^{(j-1)(p+q)}) \quad (1)$$

In general, the multiresolution approach does not offer computation complexity reduction comparatively to the complexity of the algorithm $\mathcal{A}$ applied in a single step. This is due to the presence of term $N^{p+q}$ in the polynomial function $T_{MR}$. But for some specific cases, $T_{MR}$ simplifies nicely as shown below.

From formula (1), we see that if $q \leq 0$,

$$T_{MR} = O(\rho^q N^{p+q} \rho^{r(p+q)}) + O(\rho^q N^{p+q} \sum_{j=1}^{r} \rho^{(j-1)(p+q)}) = O(\rho^q N^{p+q} \sum_{j=1}^{r+1} \rho^{(j-1)(p+q)})$$

In that case: $T_{MR} = O\left(\rho^q N^{p+q} \frac{1-\rho^{(r+1)(p+q)}}{1-\rho^{(p+q)}}\right) \leq O\left(\rho^q N^{p+q} \frac{1}{1-\rho^{(p+q)}}\right)$ if $p+q \neq 0$ $\quad (2)$
$= O(\rho^q (r+1))$ otherwise.

hence, if $p + q \neq 0$ and $q<0$, we note that $T_{MR}$ is upper bounded by a polynomial function of $N$ that does not depend upon $r$, the number of iterations required by the multiresolution algorithm. Furthermore, if $p + q > 0$ and $q < 0$, the multiresolution algorithm brings a complexity reduction of order $N^{-q}$.

As an example, for $p = 2$ and $q = -1$, we get the upper bound: $T_{MR} \leq O\left(\frac{N}{\rho(1-\rho)}\right)$

(3)

Since this upper bound is independent from *K* and *r*, we show that the time complexity of the multiresolution approach is *O(N)*. This lower bound is minimized for $\rho = 1/2$, i.e. when at each step the number of segments is halved.

It is not straightforward to generalize this result for non polynomial complexities. Nevertheless if the complexity of $\mathcal{A}$ is in *O(f(N))*, where *f* is independent from *K*, (e.g. *f(N)=NLog(N)*) one cannot expect to reduce this complexity using multiresolution.

## 2.2. Effectiveness considerations

There are two sources of error increasing in multiresolution approximation in comparison with polygonal approximation obtained in a single step:

1. In multiresolution approximation, vertices for the next level of resolution should be selected among the vertices available at the current level of resolution. This constraint does not exist for single step approximation.

2. The non-optimality of algorithm used to solve *min-ε* polygonal approximation produces error propagation inside multiresolution approximation.

We cannot reduce inaccuracies related to the first source of errors, but if we use near-to-optimal algorithm to solve *min-ε* problems between successive levels of resolution, one can expect to approach also near-to-optimal solutions for multiresolution approximation.

## 2.3 Conditions for optimizing the (*Efficiency.Effectiveness*) product

The condition to maximize the efficiency gain is to select an algorithm $\mathcal{A}$ that solves *min-ε* problems in one step with a complexity $T = O(N^p K^q)$ where $p + q > 0$ and $q < 0$: In that case, the complexity gain is $N^{-q}$.

The condition to maximize the effectiveness or accuracy of the multiresolution algorithm is to select the optimal algorithm $\mathcal{A}$ that solves optimally the *min-ε* problem.

As we know, the optimal algorithm (or the so-called *FSDP* : "Full Search Dynamic programming") is the one proposed by (Perez and Vidal, 1994): it solves the *min-ε* problem with a complexity in $O(K \cdot N^2)$ (*p=2* and *q=1*). Unfortunately, the condition $q < 0$ is not satisfied since *q=1*. In the other hand, the near-to-optimal solution developed by (Kolesnikov and Fränti, 2003), the so-called "*Reduced Search Dynamic Programming*" (*RSDP*) has a complexity in $O(N^2/K) = O(N^2 K^{-1})$ (*p=2* and *q=-1*) as far as no reference solution is used: with this restriction, *RSDP* satisfies the conditions of a complexity gain for the multiresolution solution. If *RSDP* is selected as the algorithm $\mathcal{A}$ then the complexity of the multiresolution is in *O(N)* as stated by equation (3), whatever the decimation factor $\rho$ is. Comparatively, the OSA multiresolution approache gives the coarsest approximation having *K* segments in $O(N^2/K)$ if *RSDP* is selected to process the first step (*OSA-RSDP*). We experimentally verify this claim in the next section. Note that if a reference solution is used to bootstrap *RSDP*, an extra complexity cost is added. The referred methods and associated complexities are summarized in Table 1.

|  | **Bottom-up** | **Top-down** | **Other** |
|---|---|---|---|
| **Single-step** | SPLIT, $O(K \cdot N)$ | MERGE, $O(N \cdot Log(N))$ | FSDP $O(K \cdot N^2)$<br>RSDP $O(N^2/K)$ |
| **Multi-step** | OSA-RSDP, $O(N^2/K)$ | MR-RSDP, $O(N)$ |  |

TAB. 1 – Typology of methods for polygonal curve approximation and associated complexities.

## 3. Experimentations and discussion

To evaluate the quality of suboptimal algorithms, Rosin (1997) introduced a measure known as fidelity (*F*). It measures how good (or how bad) a given suboptimal approximation is in respect to the optimal approximation in terms of the approximation error:

$F = 100 \cdot \frac{E_{\min}}{E}$, where $E_{min}$ is the approximation error of the optimal solution, and $E$ is the approximation error of the given suboptimal solution. In practice, we will identify $E_{min}$ to the error (the Euclidian distance between the original time series and the polygonal approximation) obtained using the 'Full Search' dynamic programming (*FSDP*) solution, namely the algorithm of Perez and Vidal (Perez and Vidal, 1994).

The *MR* algorithm we evaluate uses a simplified version of the *RSDP* algorithm for which the reduced search space is determined by a fixed size corridor $\beta$ (Sakoe and Chiba, 1978, Kolesnikov and Fränti, 2003, Marteau and Ménier, 2006) without using a reference solution. This version of the *RSDP* is known to have a complexity in $O(N^2/K)$ (Horng, 2002; Kolesnikov and Fränti, 2003). This ensures that the *MR* algorithm has a complexity of $O(N)$ as shown in the previous section.

The evaluation consists essentially in measuring the average fidelity and average processing time for various parameter settings (i.e. *N, K, $\beta$, $\rho$*) of the *MR* algorithm comparatively to other solutions (*SPLIT*, *MERGE*, *FSDP*). We have tested the *MR* algorithm on a dataset composed with ten 2D coastal maps extracted from the Western Europe data (FIG.1) available at the National Geophysical Data Center (NGDC, 2006).

\*\*\*\* *FIG. 1 around here* \*\*\*\*

*FIG.2* shows the evolution of the average *F* measure as a function of *K* evaluated for the crudest approximation map given by *MR (*with $\beta$=4 and $\rho$= 1/2), Merge_L2 (*MERGE*), *RSDP (*with $\beta$=4) and *Douglas Peucker (SPLIT)*.

\*\*\*\* *FIG. 2 around here* \*\*\*\*

*FIG.3* is a scatter plot that allows evaluating simultaneously the *SPLIT*, *MERGE* and *MR* methods along the average *F* and average *Processing Time* axes. Clearly, *MR* shows to be a compromise between processing speed and fidelity.

*\*\*\*\* FIG. 3 around here \*\*\*\**

*FIG.4* is a scatter plot that gives the average *F* measure against the average processing time for the *MR* method when the decimation parameter $\rho$ takes value into *{1/8, 1/4, 1/2, 3/4, 7/8}*. As theoretically expected, the processing time for *MR* is the lowest when $\rho=1/2$. Furthermore, the lower $\rho$, the better is *F*. The location of the *RSDP, FSDP, SPLIT* and *MERGE* methods are also provided.

*\*\*\*\* FIG. 4 around here \*\*\*\**

*FIG.5* compares the average processing time of the tested algorithms: i) the optimal solution (*FSDP*), ii) the "Reduced Search" dynamic procedure with fixed size corridor solution (*RSDP*), iii) the multiresolution algorithm (*MR*), iv) the Douglas-Peucker algorithm (*SPLIT*) and v) the *Merge_L2* (*MERGE*) algorithm. The average processing time is measured in micro-seconds spent while *N* increases and *K* remains unchanged (*K=10*) on a Pentium 4 processor running Linux. The scale used in *FIG. 5* is logarithmic, so that all curves are almost linear with different slopes. The figure shows that the Douglas-Peucker processing time curve has almost the same slope than the *MR* processing time curve. As the Douglas-Peucker (*SPLIT*) algorithm is known to be $O(K \cdot N)$, these two algorithms have linear processing time if *K* is maintained constant even though *MR* is more expensive, since the *MR* curve is above the *SPLIT* curve. *FSDP* and *RSDP* curves have the highest slopes, and as such exhibits a polynomial processing time as expected (*FSDP* complexity is in $O(K \cdot N^2)$ and *RSDP* with fixed size corridor is in $O(N^2 / K)$. The *MERGE* processing time curve has a slope in between *FSDP* and *SPLIT* curves as expected since *MERGE* complexity is known to be in $O(N \cdot \text{Log}(N))$.

**** FIG.5 around here ****

## 4. Conclusion

We have explored multiresolution (*MR*) applied to the problem of simplifying a discrete curve using nested polygonal approximations. It consists in iteratively applying a simplification algorithm to get the successive nested approximations, from the finest (the original curve) to the coarsest. We have shown both theoretically and practically that, when the simplification is based on a *Reduced Search Dynamic Programming* (*RSDP*) algorithm, *MR* has a linear time complexity $O(N)$, whatever the chosen number of resolution levels. This solution is suboptimal but maintains partial optimality between each resolution levels. It offers good approximating solutions when real time and storage space are issues, namely each time the optimal solution cannot be calculated due to the size of *N*. For all tests we have performed, the quality of the resulting approximation is significantly better than the quality of well known heuristic approaches (the Douglas-Peucker splitting approach or the merge approach that have respectively a complexity in $O(K \cdot N)$ and $O(N \cdot \text{Log}(N))$: the gain on the quality measure *F* varies from *30%* to *50 %* according to the tuning of parameters. As the tests suggest, *MR* associated to *RSDP* provides the best polygonal approximating algorithm having complexity in $O(N)$. The experimental results give some highlights for the choice of the parameters, i.e. $\rho$ and $\beta$, that could be tuned according to the task. Furthermore, the multiresolution aspect of the method allows managing simultaneously various resolution levels, a functionality that could be very useful to speed up time series information retrieval tasks.

# References

Buttenfield, B.P., (2002). Transmitting vector geospatial data across the Internet, *Proc. GIScience*, Berlin, Lecture Notes in Computer Science, vol. 2478, pp. 51-64.

Chan, W.S., Chin, F., (1996). On approximation of polygonal curves with minimum number of line segments or minimum error. Int. J. Comput. Geometry Appl. 6, pp. 59–77.

Charbonnier, S., Becq, G., and Biot, L., (2004). On-Line Segmentation Algorithm for Continuously Monitored Data in Intensive Care Units IEEE Trans. On Biomedical Engeneering, Vol. 51, No. 3, pp. 484 - 492

Chen, D.Z., Daescu, O., (1998). Space-efficient algorithms for approximating polygonal curves in two-dimensional space, in: Proc. of the 4th Ann. Int. Conf. on Computing and Combinatorics, Taipei, Lecture Notes in Computer Science, 1449, pp. 45–54, Springer, Berlin.

Debled-Rennesson, I., Remy, J.-L., Rouyer, J. (2003). Segmentation of discrete curves into fuzzy segments, *Electronic Notes in Discrete Mathematics*, vol. 12, pp. 372-383.

Douglas, D.H., Peucker, T.K., (1973). Algorithm for the reduction of the number of points required to represent a line or its caricature. The Canadian Cartographer 10 (2), pp.112–122.

J.-H. Horng, (2002), Improving fitting quality of polygonal approximation by using the dynamic programming technique, Pattern Recognition Letters, 23(4), pp. 1657-1673.

Imai, H., Iri, M., (1988). Polygonal approximations of a curve (formulations and algorithms). In: Toussaint, G.T. (Ed.), Computational Morphology. North-Holland, Amsterdam, pp. 71–86.

Katsaggelos, A.K., Kondi, L.P., Meier, F.W., Osterman, J., Schuster, G.M., (1998). MPEG-4 and rate-distortion-based shape-coding techniques. Proc. IEEE 86 (6), pp. 1126–1154.

Kolesnikov A. and Fränti P., (2003). Reduced-search dynamic programming for approximation of polygonal curves Pattern Recognition Letters 24, pp. 2243–2254.

Kolesnikov, A., Fränti, P., Wu, X., (2004). Multiresolution Polygonal Approximation of Digital Curves, Proceedings of the 17th International Conference on Pattern Recognition (ICPR'04), pp. 1051-4651.

Le Buhan J., C. , Ebrahimi, T., Kunt, M. , (1998). Progressive content-based shape compression for retrieval of binary images, *Computer Vision and Image Understanding*, 71(2), pp. 198-212.

Marteau, P.-F.,  Ménier, G., (2006). Adaptive Multiresolution and Dedicated Elastic Matching in Linear Time Complexity for Time Series Data Mining, Sixth International conference on Intelligent Systems Design and Applications (ISDA 2006), Jinan Shandong, China, Vol. 1, pp. 700 - 70.

Melkman, A. and O'Rourke, J. (1988). "On polygonal chain approximation", *Computational Morphology*, Elsevier Science Publishers B.V. (North Holland),  pp. 87-95.

NGDC, (2006). National Geophysical Data Center, NOAA Satellite and Information services http://www.ngdc.noaa.gov/mgg/shorelines/

Perez, J.C., Vidal, E., (1994). Optimum polygonal approximation of digitized curves, *Pattern Recognition Letters*, 15, pp. 743-750.

Pikaz, A., Dinstein, I., (1995). An algorithm for polygonal approximation based on iterative point elimination, *Pattern Recognition Letters*, 16 (6), pp. 557-563.

Pratt K.B, Fink, E. (2002). **Search for patterns in compressed time series.** International Journal of Image and Graphics, 2(1), pages 89-106.

Rosin, R.L., (1997). Techniques for assessing polygonal approximations of curves. IEEE Trans. Pattern Analysis and Machine Intelligence. 14, 659–666.

Sakoe, H., Chiba, S., (1978). Dynamic programming algorithm optimization for spoken word recognition. IEEE Trans Acoustics Speech Signal Process ASSP 26:43–49

Salotti, M., (2001); An efficient algorithm for the optimal polygonal approximation of digitized curves, *Pattern Recognition Letters*, vol. 22, pp. 215-221.

Visvalingam M., Whyatt J., (1993). Line generalization by repeated elimination of points, *Cartographic Journal*, 30(1): 46-51.

Zhu, Y., Seneviratne, L.D., (1997). Optimal polygonal approximation of digitized curves. IEEE Proc.-Vis. Image Signal Process. 144 (1), 8–14.

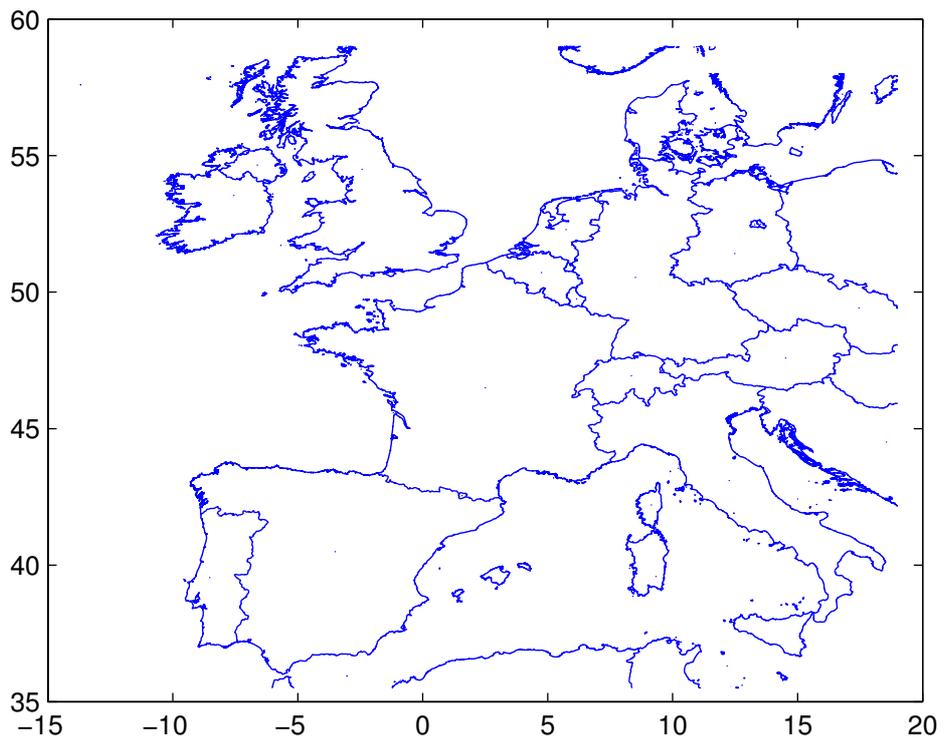

FIG. 1 –Western Europe map from which the testing dataset containing 10 disjoint 2D time series has been constructed.

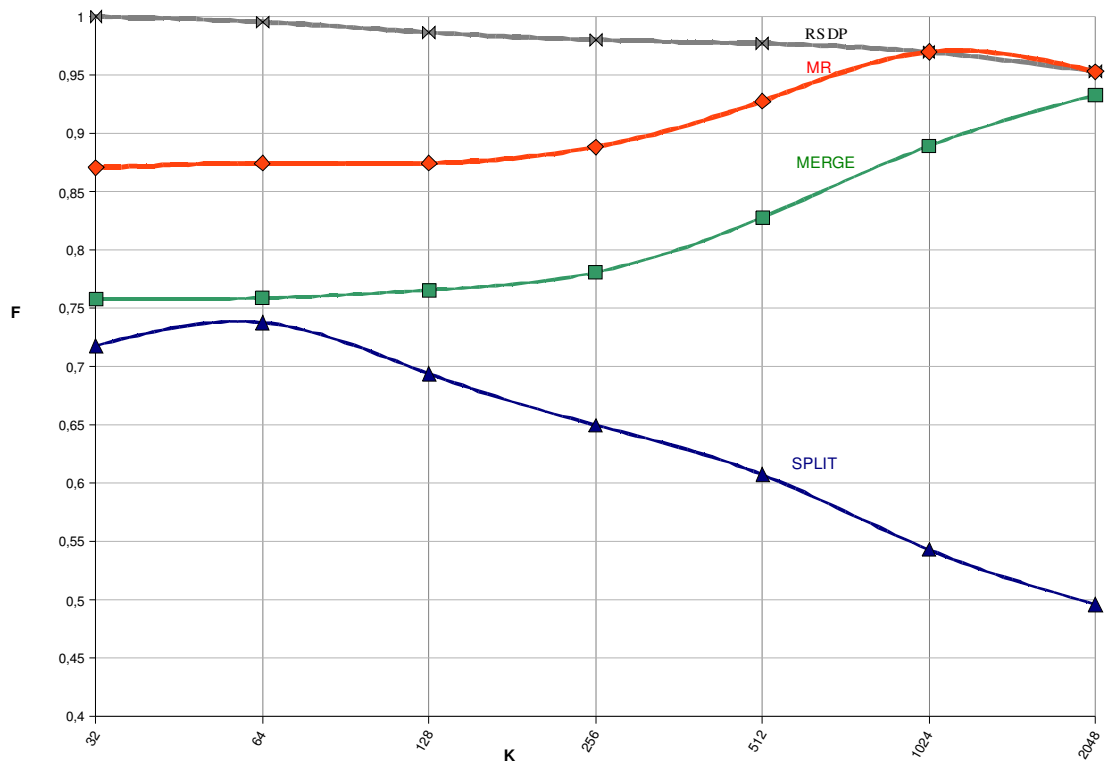

FIG. 2 – Average of the fidelity measure *(F)* as a function of *K* evaluated for the crudest approximation map given by the *MR* with *β=4* and *ρ=.5( MR*, diamond, red)*, Merge_L2* (*MERGE*, square, green) , *Douglas Peucker* (*SPLIT*, triangle, blue) and *RSDP* with fixed size corridor and *β=4* (*RSDP*, right-left triangle, grey). For all methods, *N* is constant equal to *4096*.

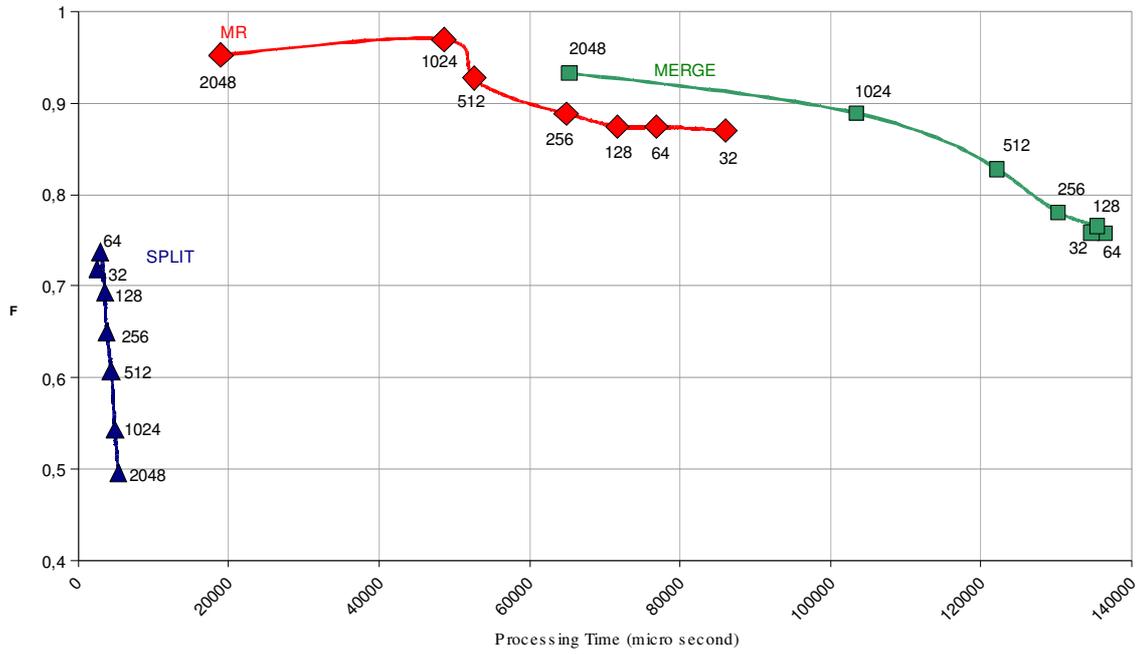

FIG. 3 – Average of the fidelity measure *(F)* as a function of the average processing time evaluated for the crudest approximation map given by the *MR* with $\beta=4$ and $\rho=.5$ ( *MR*, diamond, red), *Merge_L2* (*MERGE*, square, green) and *Douglas Peucker* (*SPLIT*, triangle, blue). Each curve is labelled with a *K* value. For all methods, *N* is constant equal to *4096*.

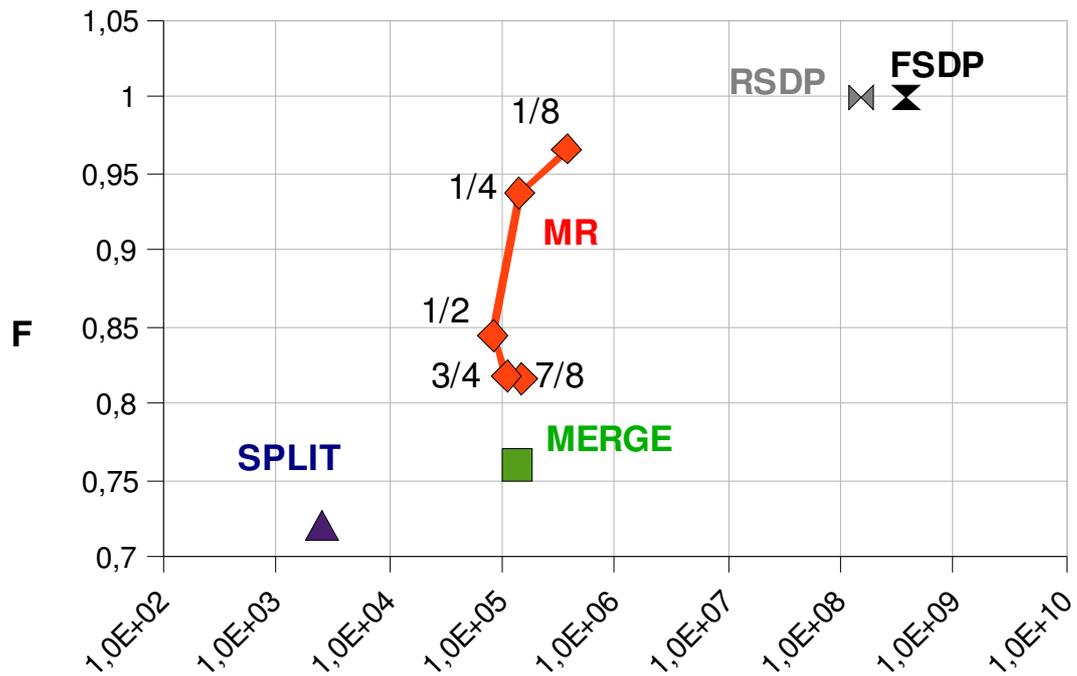

FIG. 4 – Average of the fidelity measure *(F)* as a function of the average processing time evaluated for the crudest approximation map given by the *MR* ( MR, diamond, red), for K=32, $\beta=4$ and $\rho$ in *{1/8, 1/4, 1/2, 3/4, 7/8}*. *Merge_L2* (*MERGE*, square, green), *Douglas Peucker* (*SPLIT*, triangle, blue), "*Full Search*" (*FSDP,* up-down triangle, black*)* and *RSDP* "*Reduced Search*" (*FSDP,* left-right triangle, grey*)* evaluated for *K=32* and *N=4096*.

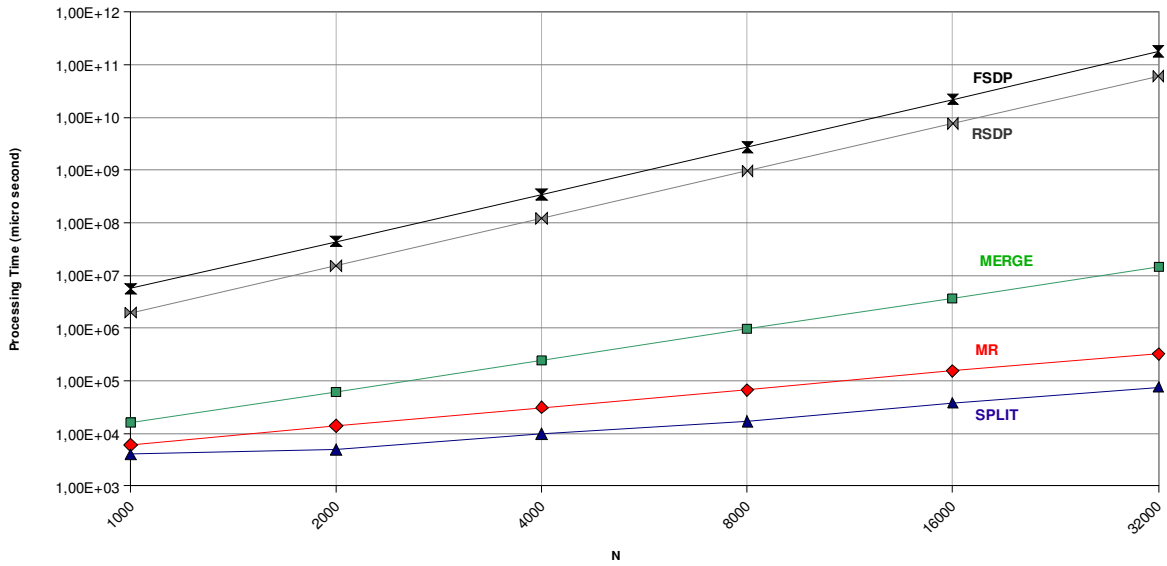

FIG. 5 –Comparison of processing time (expressed in micro seconds) on a logarithmic scale when *N* varies for *Douglas-Peucker* algorithm (*SPLIT*, triangle, blue), *Multi Resolution* algorithm (*MR*, diamond, red), *Merge_L2* (*MERGE*, square, green) and "*Full Search*" dynamic programming procedure (*FSDP*, up-down triangles, black) and "*Reduced Search*" dynamic programming procedure with fixed size corridor and *alpha=4* (*RSDP*, right-left triangle, grey). Here *K = 10* for all methods, and $\rho=.5$, $\beta=4$ for *MR* and *RSDP*.